\documentclass[jmp,twocolumn,floatfix,aps,eqsecnum,superscriptaddress]{revtex4-2}
\usepackage{graphicx}
\usepackage{amsmath}
\usepackage{amssymb}
\usepackage{mathtools}
\usepackage{bm}
\usepackage{hyperref}

\usepackage{color}

\begin{document}

\title{Monitored chaotic scattering}
\author{C. W. J. Beenakker}
\affiliation{Instituut-Lorentz, Universiteit Leiden, P.O. Box 9506, 2300 RA Leiden, The Netherlands}
\author{J. S\'{a}nchez Fern\'{a}n}
\affiliation{Instituut-Lorentz, Universiteit Leiden, P.O. Box 9506, 2300 RA Leiden, The Netherlands}
\author{J. Tworzyd{\l}o}
\affiliation{Faculty of Physics, University of Warsaw, ul.\ Pasteura 5, 02--093 Warszawa, Poland}

\date{June 2026}

\begin{abstract}
We extend the random-matrix theory of chaotic scattering to quantum dots whose dynamics is monitored by time-resolved measurements. Starting from a scattering matrix drawn from a circular ensemble, we construct the corresponding ensemble of Kraus operators for the monitored evolution of the many-body density matrix. In the single-particle sector the sum over measurement outcomes can be carried out algebraically, giving a discrete-time quantum master equation for the transferred charge. We solve this equation numerically and compare the resulting charge-transfer statistics with closed-form random-matrix predictions. The latter rely on an equipartition rule for monitored particles, which we formulate as a conjecture and test against the master equation.\medskip\\
\textit{Dedicated to Uzy Smilansky on the occasion of his 85th birthday.}
\end{abstract}
\maketitle

\section{Introduction}

In 1990 Bl\"{u}mel and Smilansky found that the statistical properties of the quantum mechanical scattering matrix of a classically chaotic system are described by the circular ensemble of random-matrix theory \cite{Blu90,Smi90}. That ensemble of random unitary matrices was introduced thirty years earlier by Dyson \cite{Dys62}, as a more tractable alternative to Wigner's ensemble of random Hermitian matrices \cite{Wig57}. Bl\"umel and Smilansky gave the first application to a physical system. This has since been developed into a full transport theory of chaotic quantum dots \cite{Bee97}.

In more recent times, the interest in quantum information processing \cite{Nielsen} has motivated the development of a framework of ``monitored quantum dynamics'', where coherent unitary evolution alternates with projective measurements \cite{Jia22,Szy23,Bul24,Ger24,Fer24,Tho24,Xia25,Gur25,Bee25,Pan25,Pic25,Ham25,Pob26,Del26,San26}. It is the purpose of this work to implement such an approach to electrical conduction through a chaotic quantum dot. The essential change is that the scattering matrix is no longer the fundamental object: Once the dynamics is monitored, the appropriate object is a quantum channel, or equivalently an operator-sum representation by Kraus operators.

Quite generally, the interplay of unitary evolution and measurements is described by a quantum channel for the density matrix $\hat{\rho}$, a completely positive map of the form \cite{Griffiths}
\begin{equation}
\hat{\rho}_{\rm final}=\sum_s \hat{K}_s\hat{\rho}_{\rm initial}\hat{K}_s^\dagger.\label{rhofinalKraus}
\end{equation}
The set of Kraus operators $\hat{K}_s$ is a resolution of the identity operator $\hat{I}$,
\begin{equation}
\sum_s \hat{K}_s^\dagger \hat{K}_s^{\vphantom{\dagger}}=\hat{I},\label{Kraussumrule}
\end{equation}
which guarantees that the map \eqref{rhofinalKraus} is trace preserving. 

Without measurements there is only a single Kraus operator
\begin{equation}
\hat{K}_0=\exp\left(i\sum_{n,m}a_n^\dagger H_{nm}a_m\right),
\end{equation}
given in terms of the fermion creation and annihilation operators $a_n^{\vphantom{\dagger}},a_n^\dagger$ of the scattered modes, with Hermitian matrix $H$ obtained from the unitary scattering matrix $S$ by $S=e^{iH}$. When measurements are introduced, multiple sub-unitary Kraus operators $\hat{K}_s$ appear, labeled by the measurement outcome $s$. 

Our goal is to construct the analogue of the circular unitary ensemble for this monitored setting: an ensemble of Kraus operators induced by circular-ensemble scattering matrices and by a fixed time-resolved measurement protocol. We may refer to this ensemble as the ``circular Kraus ensemble'', but no invariance under arbitrary unitary rotations of the Kraus representation is implied.

\begin{figure}[tb]
\centerline{\includegraphics[width=0.7\linewidth]{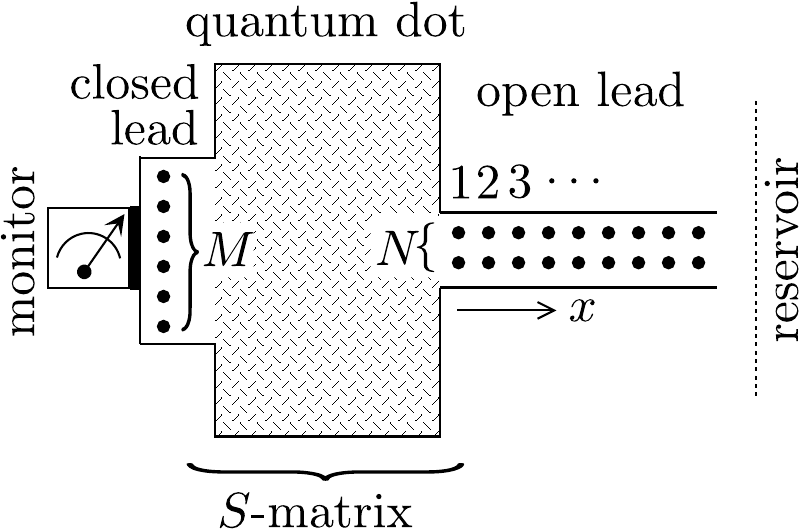}}
\caption{Quantum dot coupled to $N$ modes in an open lead and $M={\cal N}-N$ modes in a closed lead. The occupation of the $M$ modes is monitored, while the $N$ modes are connected to an electron reservoir. Chaotic scattering is modeled statistically by drawing the full ${\cal N}\times {\cal N}$ scattering matrix $S$ uniformly from the unitary group.
}
\label{fig_layout}
\end{figure}

We should state at the outset what this construction does and does not capture. All observables considered here --- the transferred charge, its distribution over the ensemble, and its counting statistics --- are linear functionals of the density matrix, so they are determined by the unconditional channel \eqref{rhofinalKraus}, in which the measurement outcomes are summed over. In the context of quantum information processing one deals with  quantities of a different type: entanglement entropies, purities, and the measurement-induced phase transitions they exhibit. These are nonlinear in $\hat{\rho}$ and require the conditional states (``quantum trajectories'') labeled by the individual outcome strings $\bm{s}$; they are not accessible from the outcome-averaged evolution and we do not consider them here. What the construction retains is the full set of $\bm{s}$-resolved Kraus operators, so the conditional information is present and could be used, for example, to compute correlations between the measurement record and the transferred charge. We leave that for future work.

The system we have in mind, see Fig.\ \ref{fig_layout}, is inspired by the closed-system variation \cite{Bee05} of B\"{u}ttiker's open-system (``voltage probe'') model of dephasing \cite{But86}, in which an $N$-mode scatterer is coupled to an additional $M$ modes. Current enters and leaves the system via $N$ modes, while the $M$ additional modes draw no current. In Ref.\ \onlinecite{Bee05} voltage fluctuations were introduced to model dephasing, here we instead work with time-resolved projective measurements of the occupation numbers of the $M$ modes.

The outline of the paper is as follows. In the next section we formulate the scattering problem at the level of the many-particle Fock space, needed to construct the Kraus operator sum. The observable we study is the transferred charge in response to a voltage bias, which we represent in terms of the density matrix map in Sec.\ \ref{sec_electricalconduction}. In Sec.\ \ref{sec_QME} we specify to the single-particle sector of Fock space, where the operator map can be reduced to a vector-space recursion relation (a ``quantum master equation''), which can be solved efficiently using numerical linear algebra methods. 

To obtain insight into the solution we turn to random-matrix theory (RMT): In Secs.\ \ref{sec_singlemode} and \ref{sec_multimode} we show that the numerical solution of the master equation can be accurately described by an RMT calculation based on a conjectured ``equipartition rule'', which says that monitored electrons are distributed equally over the outgoing modes. This closed-form solution allows for direct comparison with the voltage-probe model of dephasing in Sec.\ \ref{sec_comparison}. While most of what follows addresses the conductance of the quantum dot, the shot noise is briefly considered in Sec.\ \ref{sec_noise}. We conclude in Sec.\ \ref{sec_conclude}.

\section{Construction of the Kraus operator ensemble}

\subsection{Scattering formulation}

Referring to Fig.\ \ref{fig_layout}, we consider the scattering of a number ${\cal N}$ of fermionic modes, described by an ${\cal N}\times {\cal N}$ unitary scattering matrix $S$, drawn uniformly with the Haar measure from the unitary group ${\rm U}({\cal N})$. This is the circular unitary ensemble (CUE) of random-matrix theory, relevant for chaotic scattering in the absence of time-reversal symmetry. (We will later on consider also the time-reversally symmetric case.)

A subset of $N$ modes is connected to an electron reservoir, which may inject and absorb electrons. We will refer to these as modes in an open lead. The quantum dot is coupled to additional $M={\cal N}-N$ modes, the occupation of which is measured (monitored). The lead supporting these modes is closed at one end by an ideal reflector, without any mode mixing. In order of magnitude, $M\simeq k_{\rm F}L$, with $\hbar k_{\rm F}$ the Fermi momentum and $L$ the linear dimension of the quantum dot. Chaotic scattering requires $M\gg N$ and $M\gg 1$, while $N$ may be of order unity.

Fock space is constructed by the vector 
\begin{equation}
\bm{c}=(a_1,a_2,\ldots a_N,b_1,b_2,\ldots b_M)
\end{equation}
of ${\cal N}$ fermion operators, representing the quantum dot. These are supplemented by a discrete set of fermion operators $a^{\rm in}_n(x)$, $a^{\rm out}_n(x)$ representing incoming and outgoing modes in the open lead, with mode index $n\in\{1,2,\ldots N\}$ and site index $x\in\{1,2,3,\ldots\}$ (lattice constant $a_0$).

We adopt a stroboscopic description, with scattering events at discrete times $t\in\{1,2,3,\ldots\}$ (in units of $\delta t=a_0/v_{\rm F}$, with Fermi velocity $v_{\rm F}$). The dwell time of an electron in the quantum dot is of order $\tau_{\rm dwell}=(M/N)L/v_{\rm F}$. We set $a_0\simeq L$, so that the stroboscopic time interval $\delta t\ll \tau_{\rm dwell}$ for $M\gg N$. 

At each scattering event the matrix $S=e^{iH}$ mixes the quantum dot modes, $\bm{c}\mapsto S\bm{c}$. This may be equivalently written as a unitary conjugation,
\begin{equation}
c_p\mapsto \hat{U}^\dagger c_p \hat{U},\;\;
\hat{U}=\exp\left(i \bm{c}^\dagger H{\bm c}^{\vphantom{\dagger}}\right).
\end{equation}

After each scattering event the operator $\hat{D}$ displaces the modes in the open lead by one lattice site, towards larger $x$ for outgoing modes and towards smaller $x$ for incoming modes. This unitary operation is compactly represented by
\begin{subequations}
\label{Thatdef}
\begin{equation}
\hat{D}\psi_n(x)=\psi_n(x+1),
\end{equation}
in terms of the field $\psi_n(x)$, $n\in\{1,2,\ldots N\}$, $x\in\mathbb{Z}$, defined by
\begin{equation}
\psi_n(x)=\begin{cases}
a^{\rm out}_n(x)&\text{if}\;\;x\geq 1,\\
a^{\rm in}_n(-x)&\text{if}\;\;x\leq -1,\\
a_n&\text{if}\;\;x=0.
\end{cases}
\end{equation}
\end{subequations}

The resulting phase coherent evolution of the density matrix is
\begin{equation}
\hat{\rho}(t+1)=\hat{D}\hat{U}\hat{\rho}(t)(\hat{D}\hat{U})^\dagger\equiv \hat{\cal L}[\hat{\rho}(t)],\;\;t=0,1,2,\ldots,\label{rhotplus1coherent}
\end{equation}
with linear recursion operator $\hat{\cal L}$.

\subsection{Monitoring}

We now introduce monitoring by alternating the unitary evolution with a measurement of the occupation of each mode $m$ in the closed lead. We consider a weak measurement, interpolating with strength $w_m\in[0,1]$ between the identity $\hat{I}$ and a projection onto a filled $(+)$ or empty $(-)$ mode $m\in\{1,2,\ldots M\}$. The $2M$ measurement operators are
\begin{subequations}
\label{Ppmdef}
\begin{align}
&\hat{P}_{+,m}=\beta_m\hat{I} +(\alpha_m-\beta_m) b_{m}^\dagger b_{m}^{\vphantom{\dagger}},\\
&\hat{P}_{-,m}=\beta_m\hat{I}+(\alpha_m-\beta_m) b_{m}^{\vphantom{\dagger}}b_{m}^\dagger,\\
&\alpha_m=2^{-1/2}\sqrt{1+{w_m}},\;\;\beta_m=2^{-1/2}\sqrt{1-{w_m}}.
\end{align}
\end{subequations}
The normalization
\begin{equation}
\hat{P}_{+,m}^\dagger \hat{P}_{+,m}^{\vphantom{\dagger}}+\hat{P}_{-,m}^\dagger \hat{P}_{-,m}^{\vphantom{\dagger}}=\hat{I}\label{normalization}
\end{equation}
defines a POVM (positive operator-valued measure \cite{Griffiths}).

Measurements replace the map \eqref{rhotplus1coherent} by a sum over different measurement outcomes,
\begin{equation}
\hat{\rho}(t+1)=\sum_{\bm s}\hat{K}_{\bm s}\hat{\rho}(t)\hat{K}^\dagger_{\bm s},\label{rhotplus1inc}
\end{equation}
with Kraus operators
\begin{equation}
\hat{K}_{\bm s}= \left(\prod_{m=1}^M\hat{P}_{s_m,m}\right)\hat{D}\hat{U}\label{Khatdef}
\end{equation}
labeled by the binary string $\bm{s}=(s_1,s_2,\ldots s_M)$. Eq.\ \eqref{normalization} ensures that the Kraus operators satisfy the sum rule \eqref{Kraussumrule}.

\section{Electrical conduction}
\label{sec_electricalconduction}

\subsection{Transferred charge}

A voltage bias $V$ injects electrons in a subset $n\in\{1,2,\ldots N_1\}$ of the modes in the open lead (incoming modes from the source reservoir), incident on the quantum dot in an energy range $eV$ above the Fermi energy $E_{\rm F}$. We seek the charge transferred into the other set of $N_2=N-N_1$ modes (outgoing modes into the drain reservoir). For $eV\ll\hbar/\tau_{\rm dwell}$ we may neglect the energy dependence of the scattering matrix, evaluating it at the Fermi level.

As initial condition at $t=0$ we occupy each of the $N_1$ modes by one electron, spread out over a length $L_0\simeq \hbar v_{\rm F}/eV$ in the open lead. We will later implement the limit $V\rightarrow 0$ by taking the limit $L_0\rightarrow\infty$. The corresponding initial density matrix $\hat{\rho}(0)$ is a pure state, 
\begin{equation}
\begin{split}
&\hat{\rho}(0)=|\psi_0\rangle\langle \psi_0|,\\
&|\psi_0\rangle=L_0^{-N_1/2}\left[\prod_{n=1}^{N_1}\left(\sum_{x=1}^{L_0}a^{\rm in}_n(x)^\dagger\right)\right]|\emptyset\rangle,
\end{split}
\label{hatrho0}
\end{equation}
with $|\emptyset\rangle$ the unperturbed Fermi sea.

The number of electrons transferred from source to drain is counted by the operator
\begin{equation}
\hat{Q}=\sum_{n=N_{1}+1}^N \sum_{x=1}^\infty a^{\rm out}_n(x)^\dagger a^{\rm out}_n(x)^{\vphantom{\dagger}}.\label{Qdef}
\end{equation}
The total transferred charge ${\cal T}$ (in units of the electron charge $e$) then follows from
\begin{align}
{\cal T}={}&\lim_{t\rightarrow\infty}\operatorname{Tr}\hat{Q}\hat{\rho}(t)\nonumber\\
={}&\sum_{n=N_{1}+1}^N \sum_{t=0}^\infty\operatorname{Tr}a^{\rm out}_n(1)^\dagger a^{\rm out}_n(1)^{\vphantom{\dagger}}\hat{\rho}(t).\label{calTdef}
\end{align}
In the second equality we used the free propagation in the open lead, so that to count the total transferred charge at time $t$ we may either sum over all positions $x\geq 1$ at time $t$ or sum over all times $\leq t$ at $x=1$. 

In terms of the recursion operator $\hat{\cal L}$ defined in Eq.\ \eqref{rhotplus1coherent} we have the geometric series
\begin{align}
\sum_{t=0}^\infty \hat{\rho}(t)={}&\hat{\rho}(0)+\hat{\cal L}[\hat{\rho}(0)]+\hat{\cal L}^2[\hat{\rho}(0)]+\hat{\cal L}^3[\hat{\rho}(0)]\cdots\nonumber\\
={}&(1-\hat{\cal L})^{-1}[\hat{\rho}(0)],
\end{align}
hence
\begin{equation}
{\cal T}= \sum_{n=N_{1}+1}^N \operatorname{Tr}a^{\rm out}_n(1)^\dagger a^{\rm out}_n(1)^{\vphantom{\dagger}}(1-\hat{\cal L})^{-1}[\hat{\rho}(0)].\label{hatGdef}
\end{equation}

\subsection{Differential conductance}

\begin{figure}[tb]
\centerline{\includegraphics[width=0.5\linewidth]{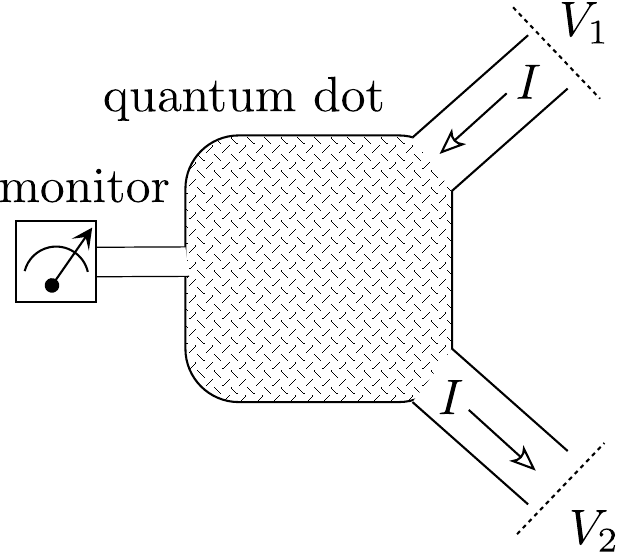}}
\caption{Two-terminal transport configuration in a quantum dot. A voltage difference $V=V_1-V_2$ between contacts 1 and 2 drives a current $I$ through the system,  corresponding to a differential conductance $G=dI/dV$. We seek the effect on $G$ of a monitoring of the scattering dynamics. The monitoring breaks reciprocity, leading to a difference between the case of an asymmetric bias $V_1=V$, $V_2=0$ and the case of a symmetric bias $V_1=V/2$, $V_2=-V/2$.
}
\label{fig_diagram}
\end{figure}

Monitoring brings the system locally out of equilibrium. As explained by Ferreira \textit{et al.} \cite{Fer24}, this has two related consequences for any two-terminal transport measurement, such as the quantum dot configuration in Fig.\ \ref{fig_diagram}:
\begin{itemize}
\item Firstly, a current $I$ may flow even in the absence of a voltage bias. The differential conductance $G=dI/dV=(e^2/h){\cal T}$ eliminates the zero-bias offset. 
\item Secondly, the current depends separately on the voltages $V_1$ and $V_2$ applied to the source and drain contacts, rather than only on the difference $V=V_1-V_2$.
\end{itemize}

Eq.\ \eqref{hatGdef} gives the transferred charge ${\cal T}$ for $V_1=V$, $V_2=0$. If we interchange source and drain contacts (injecting in modes $n=N_1+1,N_1+2,\ldots N$, and then measuring the charge transferred in modes $n=1,2,\ldots N_1$), we obtain a transferred charge ${\cal T}'$ which may differ from ${\cal T}$: the monitoring breaks reciprocity \cite{Fer24}. One may also apply the bias symmetrically, $V_1=V/2$, $V_2=-V/2$, in which case the transferred charge is given by $\overline{\cal T}=\tfrac{1}{2}({\cal T}+{\cal T}')$.

We emphasise that the breaking of reciprocity is not related to the presence or absence of time reversal symmetry, it is sufficient if the monitoring probability differs for electrons injected via contact 1 or contact 2. Statistically, in a chaotic quantum dot, these two probabilities are the same, but for an individual system they may differ.

In what follows we will consider the statistics of both ${\cal T}$ and $\overline{\cal T}$, corresponding to the differential conductance for a bias voltage that is applied either asymmetrically on a single contact, or symmetrically over both contacts.

\section{Quantum master equation}
\label{sec_QME}

\subsection{From Fock space to vector space}

For non-interacting electrons the expectation value of the bilinear charge operator \eqref{Qdef} is determined by the single-particle sector of Fock space. We may therefore work with matrices in vector space (first quantization), rather than with creation and annihilation operators in Fock space (second quantization). We notationally distinguish matrices from operators by removing the caret on the operators.

The basis vectors are $|n\rangle$, $n\in\{1,2,\ldots {\cal N}\}$ for the complex amplitudes of the ${\cal N}=N+M$ modes in the quantum dot, coupled by the ${\cal N}\times {\cal N}$ unitary matrix $S$. The complex amplitudes of incoming and outgoing modes in the open leads are $|n,x\rangle_{\rm in}$ and $|n,x\rangle_{\rm out}$, respectively, with $n\in\{1,2,\ldots N\}$ and $x\in\{1,2,3,\ldots\}$. These are displaced inward or outward by the unitary operator $D$,
\begin{subequations}
\begin{align}
&\left.
\begin{aligned}
D|n,x+1\rangle_{\rm in}=|n,x\rangle_{\rm in}\\
D|n,x\rangle_{\rm out}=|n,x+1\rangle_{\rm out}
\end{aligned}
\right\}
\;\;\text{for}\;\;x\geq 1,\\
&D|n,1\rangle_{\rm in}=|n\rangle,\;\;D|n\rangle=|n,1\rangle_{\rm out},
\end{align}
\end{subequations}
in accord with Eq.\ \eqref{Thatdef}.

The measurement operators $\hat{P}_{\pm,m}$ in Fock space map onto the vector space operators
\begin{equation}
\begin{split}
&{P}_{+,m}= \alpha_m|m\rangle\langle m|+\beta_m(1-|m\rangle\langle m|),\\
&{P}_{-,m}= \beta_m|m\rangle\langle m|+\alpha_m(1-|m\rangle\langle m|),
\end{split}
\end{equation}
for each of the $M$ monitored modes $m\in\{N+1,N+2,\ldots {\cal N}\}$ .

Combining these expressions, the single-particle density matrix evolves as
\begin{equation}
{\rho}(t+1)=\sum_{\bm s}{K}_{\bm s}{\rho}(t){K}^\dagger_{\bm s},\label{rhotsingleparticle}
\end{equation}
with Kraus matrices
\begin{equation}
{K}_{\bm s}= \left(\prod_{m=N+1}^{\cal N}{P}_{s_m,m}\right){D}S.\label{Kraus_s}
\end{equation}

As shown in App.\ \ref{app_mastereq}, the sum over the measurement outcomes $\bm{s}=\{s_{N+1},s_{N+2},\ldots s_{\cal N}\}$ can be carried out by purely algebraic means, resulting in the quantum master equation
\begin{equation}
\begin{split}
\rho(t+1)={}&QDS\rho(t)(QDS)^\dagger\\
&+\sum_{m=N+1}^{\cal N}w_m^2|m\rangle\langle m|DS\rho(t)(DS)^\dagger|m\rangle\langle m|,\\
Q={}&I+\sum_{m=N+1}^{\cal N} (\sqrt{1-w_m^2}-1)|m\rangle\langle m|,
\end{split}\label{mastereq}
\end{equation}
with $I$ the identity operator in the single-particle vector space. Compared to Eq.\ \eqref{rhotsingleparticle} the number of terms has been reduced from $2^{\cal N}$ to ${\cal N}$. We denote the corresponding linear operator by ${\cal L}$,
\begin{equation}
\rho(t+1)={\cal L}[\rho(t)].
\end{equation}

As single-particle initial condition we take
\begin{equation}
\rho_n(0)=L_0^{-1}\sum_{x,x'=1}^{L_0}|n,x\rangle_{\rm in}\langle n,x'|,\label{rho0def}
\end{equation}
to obtain the transmission probability ${T}_{n'n}$ from mode $n$ to mode $n'$,
\begin{equation}
{T}_{n'n}=\operatorname{Tr}| n',1\rangle_{\rm out}\langle n',1| (1-{\cal L})^{-1}[{\rho}_n(0)].\label{Tndef}
\end{equation}
The total transferred charge ${\cal T}$ then follows upon summation over modes $n$ in the source and modes $n'$ in the drain,
\begin{equation}
{\cal T}=\sum_{n=1}^{N_1}\sum_{n'=N_1+1}^N{T}_{n'n}.
\end{equation}
These equations are the first-quantized counterparts to Eqs.\ \eqref{hatrho0} and \eqref{hatGdef}.

\subsection{Elimination of the degrees of freedom in the open lead}

The single-particle vector space is infinite dimensional because of the sites in the semi-infinite open lead. Following Ref.\ \onlinecite{San26}, these can be eliminated algebraically, retaining only site $x=1$.

As a first step, we note that the projection of the outgoing modes onto site $x=1$ in Eq.\ \eqref{Tndef} allows us to restrict the states $|n,x\rangle_{\rm out}$ to $x=1$, for all $n$. To achieve a similar restriction of the incoming modes we decompose Eq.\ \eqref{Tndef} for the transmission probability into
\begin{equation}
{T}_{n'n}(x,x')=\operatorname{Tr}| n',1\rangle_{\rm out}\langle n',1| (1-{\cal L})^{-1}\bigl[|n,x\rangle_{\rm in}\langle n,x'|\bigr],\\
\end{equation}
so that
\begin{equation}
{T}_{n'n}=L_0^{-1}\sum_{x,x'=1}^{L_0}{T}_{n'n}(x,x').
\end{equation}

Because of translational invariance in the leads ${T}_{n'n}(x,x')={T}_{n'n}(\delta x)$ depends only on the difference $\delta x=x'-x$ of the site indices. We thus have
\begin{equation}
{T}_{n'n}=\sum_{\delta x=1-L_0}^{L_0-1}\frac{L_0-|\delta x|}{L_0}{T}_{n'n}(\delta x).
\end{equation}
The range of $\delta x$ that contributes is of order $\tau_{\rm dwell}/\delta t\simeq M/N$. In the limit $L_0\rightarrow\infty$ at fixed $M,N$ we obtain
\begin{align}
\lim_{L_0\rightarrow\infty}{T}_{n'n}={}&\sum_{\delta x=-\infty}^{\infty}{T}_{n'n}(\delta x)\nonumber\\
={}&{T}_{n'n}(0)+2\operatorname{Re}\sum_{\delta x=1}^\infty{T}_{n'n}(\delta x),\label{calTdeltan3}
\end{align}
since ${T}_{n'n}(-\delta x)={T}_{n'n}(\delta x)^\ast$.

The state $|n,1\rangle_{\rm in}\langle n,1+\delta x|$ evolves after $t=\delta x$ time steps into $(QDS)^{\delta x}|n,1\rangle_{\rm in}\langle n,1|$, without contributing to the transferred charge, hence
\begin{align}
{T}_{n'n}(\delta x)={}&\operatorname{Tr}| n',1\rangle_{\rm out}\langle n',1|\nonumber\\
&\times (1-{\cal L})^{-1}\bigl[(QDS)^{\delta x}|n,1\rangle_{\rm in}\langle n,1|\bigr].
\end{align}
Substitution into Eq.\ \eqref{calTdeltan3} gives, upon summation of the geometric series in $\delta x$,
\begin{equation}
\begin{split}
{T}_{n'n}={}&\operatorname{Re}\operatorname{Tr}| n',1\rangle_{\rm out}\langle n',1| (1-{\cal L})^{-1}\bigl[{\cal S}|n,1\rangle_{\rm in}\langle n,1|\bigr],\\
{\cal S}={}&(1+QDS)(1-QDS)^{-1}.
\end{split}\label{Tnnfinal}
\end{equation}
This expression for the transmission probability only involves site $x=1$, eliminating the semi-infinite open lead.

\subsection{Numerical evaluation}

Numerically, Eq.\ \eqref{Tnnfinal} may be evaluated by two alternative methods of linear algebra. The simplest to implement is vectorization: an ${\cal N}\times {\cal N}$ matrix $M$ is converted into a vector $\operatorname{vec}(M)$ of length ${\cal N}^2$ by stacking the columns of the matrix on top of one another. The identity 
\begin{equation}
\operatorname{vec}(ABC) = (C^\mathrm{T}\otimes A) \operatorname{vec}(B)
\end{equation}
then represents the recursion operator ${\cal L}$ by a matrix multiplication, so that $(1-{\cal L})^{-1}$ follows upon matrix inversion. Because the size of the matrices is ${\cal N}^2\times {\cal N}^2$, this approach scales poorly as ${\cal N}^6$ --- at least if the matrices are not sparse \cite{note1}.

A more efficient approach is to recognize the master equation \eqref{mastereq} as a generalized Lyapunov equation, which can be solved by Schur decomposition at ${\cal N}^3$ cost. We describe this approach in App.\ \ref{app_Lyapunov}.

\section{Single-mode monitoring}
\label{sec_singlemode}

Before turning to the results we describe the two calculations that will be compared, and what the comparison establishes. Both are based on the circular ensemble; they differ in what is done with it.

In the \textit{master equation calculation} we draw an ${\cal N}\times{\cal N}$ matrix $S$ from the CUE or COE, with ${\cal N}=M+N$ large but finite, and solve the quantum channel \eqref{mastereq} exactly by numerical linear algebra. Apart from the finite value of $M$ and the statistical error of the sampling, no approximation enters.

In the \textit{RMT calculation} we perform the limit $M\rightarrow\infty$ analytically, in two steps. First we eliminate the $M-1$ unmonitored modes of the closed lead, which reduces $S$ to an effective $3\times 3$ unitary matrix $U$, Eq.\ \eqref{Udef}, that is again distributed according to the circular ensemble. This step is exact. Second, we replace the many-step monitored dynamics of the remaining mode by the equipartition rule \eqref{conjecture}. This step is a conjecture.

The comparison in Figs.\ \ref{fig_CUE} and \ref{fig_COE} is therefore not a comparison of two independent theories, but a test of the equipartition rule: the master equation supplies the finite-$M$ answer, and the closed-form distributions show what that answer should converge to if the conjecture holds.

\subsection{Master equation calculation}

\begin{figure}[tb]
\centerline{\includegraphics[width=1\linewidth]{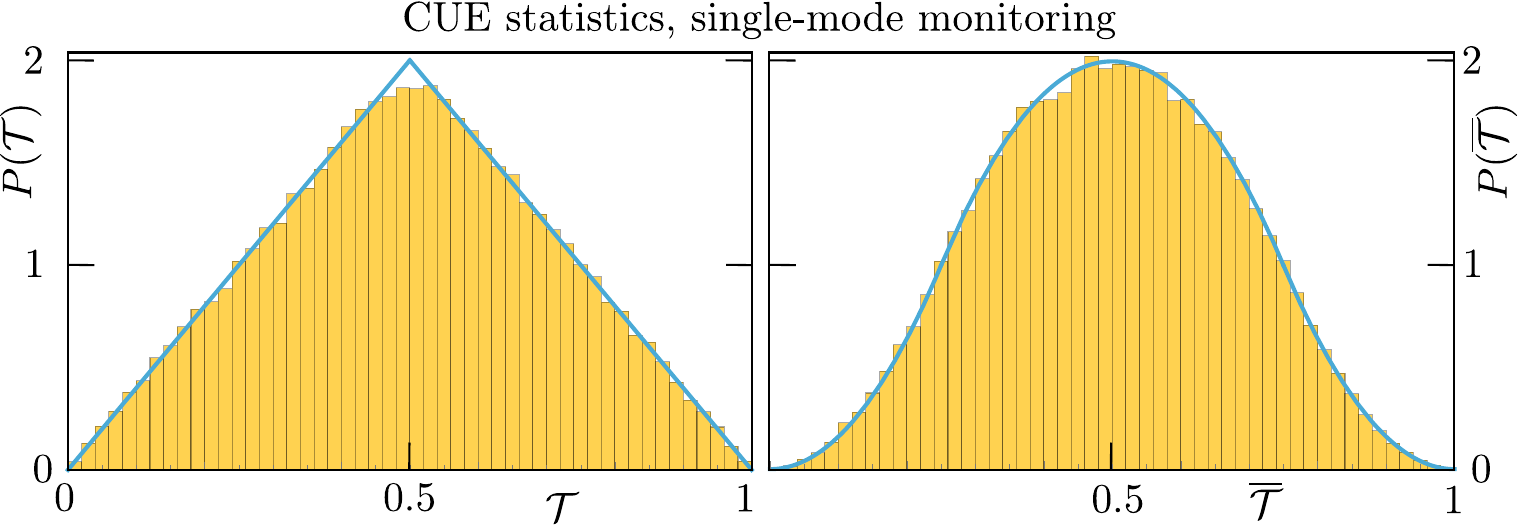}}
\caption{Charge transfer statistics for single-mode monitoring. The two panels show the probability density functions of the transferred charge ${\cal T}$ (from mode 1 to mode 2) and $\overline{\cal T}$ (the average of the charge transfer from mode 1 to mode 2 and from mode 2 to mode 1). The histograms are computed numerically from the master equation \eqref{monitoredmastereq} for $M=100$, the solid curves are derived from the CUE in the limit $M\rightarrow\infty$, assuming Eq.\ \eqref{conjecture}.
}
\label{fig_CUE}
\end{figure}

We focus on the case that the occupation of one single mode out of the $M$ modes in the closed lead is monitored. The other $M-1$ modes are not measured. In the open lead we take $N=2$ modes (single-mode source and drain). We label mode 3 as the monitored mode and take for simplicity an ideal measurement, $w_n=\delta_{n,3}$. The master equation \eqref{mastereq} reduces to
\begin{equation}
\begin{split}
&\rho(t+1)=Q{D}{S}\rho(t)(Q{D}{S})^\dagger +P{D}{S}\rho(t)(P{D}{S})^\dagger ,\\
&P=|3\rangle\langle 3|=I-Q.
\end{split}\label{monitoredmastereq}
\end{equation}

The transferred charge ${\cal T}$ from mode 1 to mode 2 is given by Eq.\ \eqref{Tnnfinal},
\begin{equation}
\begin{split}
{\cal T}={}&\operatorname{Re}\operatorname{Tr}\chi_{\rm out} (1-{\cal L})^{-1}\bigl[{\cal S}\chi_{\rm in}\bigr],\\
\chi_{\rm in}={}&|n=1,x=1\rangle_{\rm in}\langle n=1,x=1|,\\
\chi_{\rm out}={}&| n=2,x=1\rangle_{\rm out}\langle n=2,x=1|.
\end{split}\label{calTsinglemode}
\end{equation}
For ${\cal T}'$, charge transfer from mode 2 to mode 1, we exchange $n=1$ with $n=2$. These give the differential conductance for an asymmetrically applied bias voltage ($V_1=V$, $V_2=0$ or the other way around, the two cases are statistically equivalent). For the symmetrically biased case ($V_1=V/2$, $V_2=-V/2$) we have $\overline{\cal T}=\tfrac{1}{2}(\cal{T}+{\cal T}')$.

The corresponding probability density functions $P({\cal T})$ and $P(\overline{\cal T})$ are shown in Fig.\ \ref{fig_CUE} (histograms), computed from the master equation \eqref{monitoredmastereq} for $M=100$, ${\cal N}=102$, sampled over $10^5$ scattering matrices $S$ of size ${\cal N}\times {\cal N}$ in the CUE. We used the Lyapunov method of solution explained in App.\ \ref{app_Lyapunov}.

\subsection{RMT calculation}
\label{singlemodeRMTCUE}

The curves in Fig.\ \ref{fig_CUE} are calculated as follows, in the framework of random-matrix theory (RMT). Singling out modes $n\in\{1,2,3\}$ from the total number of ${\cal N}=M+2$ modes, we partition the ${\cal N}\times {\cal N}$ scattering matrix $S$ into subblocks,
\begin{equation}
S=\begin{pmatrix}
r&t\\
t'&r'
\end{pmatrix},
\end{equation}
where $r$ is $3\times 3$, $r'$ is $(M-1)\times(M-1)$, $t$ is $3\times(M-1)$ and $t'$ is $(M-1)\times 3$.

An electron injected in mode 1 propagates coherently until it is scattered into modes 1, 2, or 3. The scattering amplitudes $U_{11}$, $U_{21}$, $U_{31}$ are elements of a $3\times 3$ unitary matrix $U$ that is constructed from $S$ by
\begin{equation}
U=r+t (1-r')^{-1}t'.\label{Udef}
\end{equation}
If $S$ is uniformly distributed in ${\rm U}({\cal N})$ then $U$ is uniformly distributed in ${\rm U}(3)$ \cite{note2}.
Our calculation is based on the following\medskip\\
{\bf Conjecture I} [\textit{Single-mode equipartition rule}]:
\begin{equation}
\begin{rcases}
{\cal T}\rightarrow\tfrac{1}{2}|U_{31}|^2+|U_{21}|^2\\
{\cal T}'\rightarrow\tfrac{1}{2}|U_{32}|^2+|U_{12}|^2
\end{rcases}
\;\;\text{as}\;\;M\rightarrow\infty, \label{conjecture}
\end{equation}
for all $S$, up to a set of measure zero. This conjecture expresses the intuition that monitoring suppresses quantum fluctuations of the transmission probabilities in the limit $M\rightarrow\infty$, so that an electron scattered into mode 3 is transferred with the classical probability 1/2 into modes 1 and 2. Check that current is conserved: with ${\cal R}=\tfrac{1}{2}|U_{31}|^2+|U_{11}|^2$ one has ${\cal R}+{\cal T}=1$ identically.

Reciprocity is broken by Eq.\ \eqref{conjecture}: ${\cal T}\neq{\cal T}'$ in general, even in the presence of time reversal symmetry, which constrains $U_{21}=U_{12}$ but does not relate $U_{31}$ and $U_{32}$.

In support of the conjecture, we calculate the probability density functions $P({\cal T})$ and $P(\overline{\cal T})$ that Eq.\ \eqref{conjecture} implies, given the CUE distribution of $U$, and compare with the numerical solution of the master equation. We find (see App.\ \ref{app_CUECOE})
\begin{align}
&P({\cal T})=\begin{cases}
4{\cal T}&\text{if}\;\;0<{\cal T}<1/2,\\
4(1-{\cal T})&\text{if}\;\;1/2<{\cal T}<1,
\end{cases}\label{PTCUE}\\
&P(\overline{\cal T})=\begin{cases}
16\,\overline{\cal T}^2&\text{if}\;\;0<\overline{\cal T}<1/4,\\
2-16(\overline{\cal T}-1/2)^2&\text{if}\;\;1/4<\overline{\cal T}<3/4,\\
16(1-\overline{\cal T})^2&\text{if}\;\;3/4<\overline{\cal T}<1,
\end{cases}\label{PTbarCUE}
\end{align}
in good agreement with the numerics of Fig. \ref{fig_CUE}.

\begin{figure}[tb]
\centerline{\includegraphics[width=1\linewidth]{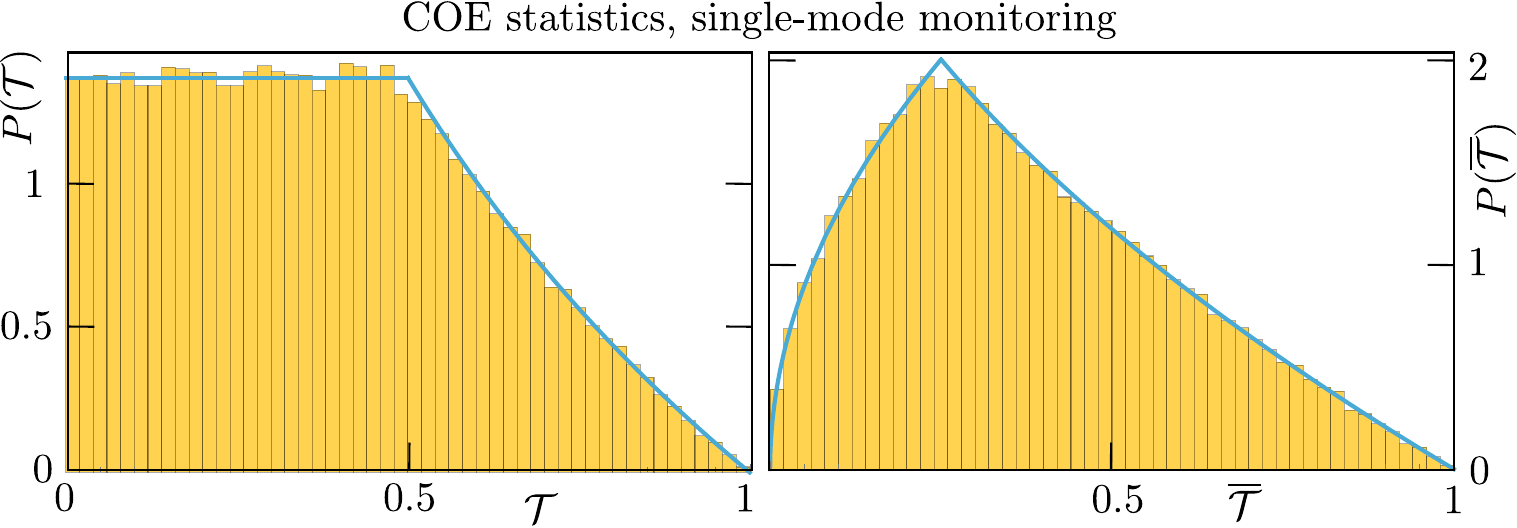}}
\caption{Same as Fig.\ \ref{fig_CUE}, but for unitary symmetric $S$ drawn from the COE.
}
\label{fig_COE}
\end{figure}

\subsection{Preserved time-reversal symmetry}
\label{singlemodeRMTCOE}

So far we assumed broken time reversal symmetry in the quantum dot. If time reversal symmetry is preserved, the scattering matrix $S\in{\rm U}({\cal N})/{\rm O}({\cal N})$ is unitary and symmetric, drawn from the circular orthogonal ensemble (COE). The numerical solution of the master equation gives the histograms shown in Fig.\ \ref{fig_COE}. If we start from the same conjectured Eq.\ \eqref{conjecture} we obtain (see App.\ \ref{app_CUECOE})
\begin{align}
&P({\cal T})=\begin{cases}
2\ln 2&\text{if}\;\;0<{\cal T}<1/2,\\
-2\ln{\cal T}&\text{if}\;\;1/2<{\cal T}<1,
\end{cases}\label{PTCOE}\\
&P(\overline{\cal T})=\begin{cases}
4\surd\,\overline{\cal T}&\text{if}\;\;0<\overline{\cal T}<1/4,\\
4(1-\surd\,\overline{\cal T})&\text{if}\;\;1/4<\overline{\cal T}<1,
\end{cases}\label{PTbarCOE}
\end{align}
again in good agreement with the numerics of Fig.\ \ref{fig_COE}.

\section{Multi-mode monitoring}
\label{sec_multimode}

\subsection{Master equation calculation}

If we contrast local and nonlocal measurements, we have the single-mode measurement of the previous section at the local extreme. At the nonlocal extreme we can consider a spatially uniform measurement, when a large number $M\gg 1$ of modes in the closed lead monitors the scattering with a small measurement strength $w\ll 1$. For single-mode source and drain we have $N=2$, ${\cal N}=M+2$, with $w_1=w_2=0$, $w_{m}=w$ for $m=3,4,\ldots {\cal N}$. We consider both the cases of unbroken and broken time-reversal symmetry ($\beta=1$ and $\beta=2$, respectively), and in each case compare the transferred charge for asymmetric and symmetric bias (${\cal T}$ and $\overline{\cal T}$, respectively).

\begin{figure}[tb]
\centerline{\includegraphics[width=1\linewidth]{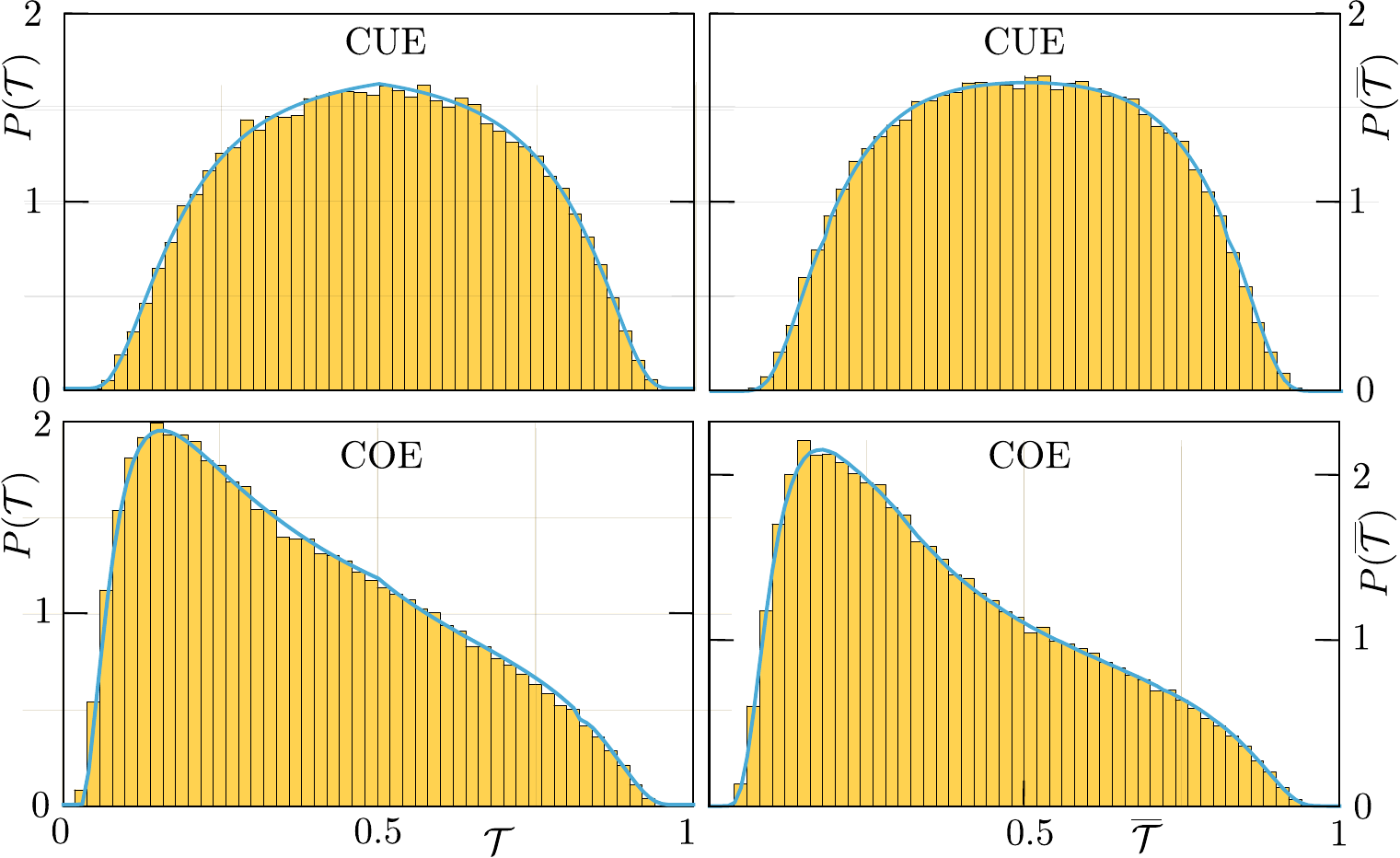}}
\caption{Charge transfer statistics for multi-mode monitoring. The panels show the probability density functions of the transferred charge ${\cal T}$ and $\overline{\cal T}$, for CUE and COE statistics of the scattering matrix. The histograms are computed numerically from the master equation \eqref{mastereq} for $M=100$, $w=0.1\Rightarrow \gamma=Mw^2=1$, the solid curves are the RMT calculations based on Eq.\ \eqref{conjecture2}.
}
\label{fig_histograms}
\end{figure}

The probability density functions $P({\cal T})$ and $P(\overline{\cal T})$ are shown in Fig.\ \ref{fig_histograms} (histograms), computed from the master equation \eqref{mastereq} for $M=100$, $w=0.1$. We again use the Lyapunov method of solution (see App.\ \ref{app_Lyapunov}).

\subsection{RMT  calculation}
\label{multimodeRMT}

We compare the solution of the master equation with an RMT calculation. To account for a non-ideal measurement ($w<1$) we insert a tunnel barrier with mode independent transmission probability $w^2$ in the closed lead. The bare scattering matrix $S$ is then replaced by
\begin{subequations}
\begin{align}
&\tilde{S}=P_w+  Q_w S\bigl(I + P_w S\bigr)^{-1} Q_w,\\
&P_w=\sqrt{1-w^2}\sum_{m=N+1}^{\cal N}|m\rangle\langle m|,\\
&Q_w=\sum_{n=1}^N|n\rangle\langle n|+w\sum_{m=N+1}^{\cal N}|m\rangle\langle m|.
\end{align}
\label{tildeSdef}
\end{subequations}
(Note that $P_w^2+Q_w^2=I$.) The circular ensemble for $S$ produces for $\tilde{S}$ the Poisson kernel distribution \cite{Bee97}.

We proceed analogously to Eq.\ \eqref{conjecture},\medskip\\
{\bf Conjecture II} [\textit{Multi-mode equipartition rule}]:
\begin{equation}
\begin{rcases}
{\cal T}=\tfrac{1}{2}\sum_{m=3}^{\cal N}|\tilde{S}_{m1}|^2+|\tilde{S}_{21}|^2\\
{\cal T}'=\tfrac{1}{2}\sum_{m=3}^{\cal N}|\tilde{S}_{m2}|^2+|\tilde{S}_{12}|^2
\end{rcases}
\begin{matrix}
\;\;\text{as}\;\;M\rightarrow\infty,\;\;w\rightarrow 0,\\
\text{at fixed}\;\;\gamma=Mw^2,
\end{matrix}
\label{conjecture2}
\end{equation}
for all $S$, up to a set of measure zero.

In App.\ \ref{app_Poisson} we calculate the probability density functions $P({\cal T})$ and $P(\overline{\cal T})$ that Eq.\ \eqref{conjecture2} implies. As shown in Fig.\ \ref{fig_histograms}, for $\gamma=1$, they agree well with the master equation.

\subsection{Reciprocity breaking}
\label{sec_reciprocity}

\begin{figure}[tb]
\centerline{\includegraphics[width=0.7\linewidth]{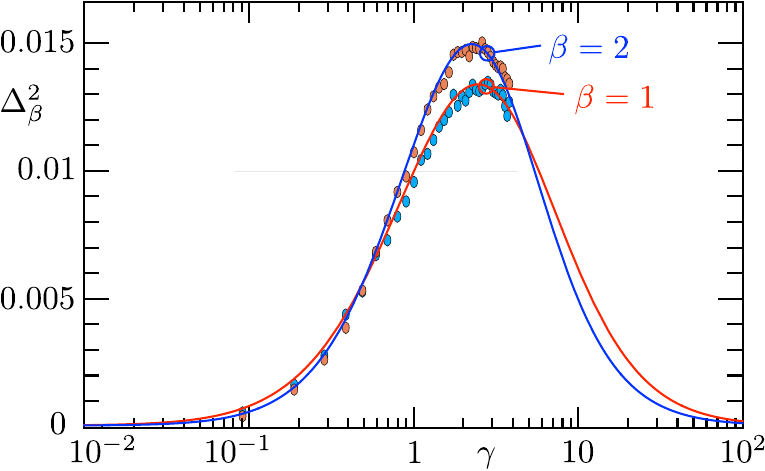}}
\caption{Reciprocity breaking by monitoring, quantified through the mean squared difference $\Delta^2_\beta$ of the transferred charge from contact 1 to contact 2 and the other way around. The solid curves are the RMT results \eqref{Delta2def}. The data points are computed numerically from the master equation \eqref{mastereq} as a function of $\gamma=Mw^2$, by varying $M$ at fixed $w=0.1$. (Only the range $\gamma\lesssim 4$ was accessible numerically.)
}
\label{fig_reciprocity}
\end{figure}

To quantify the breaking of reciprocity, we have computed the mean squared difference
\begin{equation}
\Delta^2_\beta=\mathbb{E}[({\cal T}-{\cal T}')^2].
\end{equation}
Results from the master equation are plotted in Fig.\ \ref{fig_reciprocity}. The reciprocity breaking is maximal near $\gamma=2$, vanishing both for much smaller and much larger measurement strengths.

We can compare with the RMT calculation (see App.\ \ref{app_Poisson}), which gives
\begin{widetext}
\begin{subequations}
\label{Delta2def}
\begin{align}
\Delta^2_{\beta=1}
={}& \frac{1}{12\gamma}\Bigl\{
\bigl[\gamma^3+2\gamma^2-10\gamma+16-(6\gamma+16)e^{-\gamma}\bigr]
e^{\gamma/2}\operatorname{Ei}(-\gamma/2)
\nonumber\\
&\quad
+4\bigl[\gamma^3+4\gamma^2+8\gamma+8-8e^{\gamma}\bigr]
\operatorname{Ei}(-\gamma)
+2(\gamma^2-6)+4e^{-\gamma}(\gamma^2+3\gamma+3)
\Bigr\},
\\
\Delta^2_{\beta=2}
={}&\frac{1}{2\gamma}
\bigl[\gamma-2+(\gamma+2)e^{-\gamma}\bigr]
\bigl[\gamma+1+\gamma(\gamma+2)e^{\gamma}\operatorname{Ei}(-\gamma)\bigr],
\end{align}
\end{subequations}
\end{widetext}
with Ei the exponential integral function. Fig.\ \ref{fig_reciprocity} shows good agreement with the numerical data from the master equation.

This is for multi-mode monitoring. For single-mode monitoring (the case considered in Sec.\ \ref{sec_singlemode} and App.\ \ref{app_CUECOE}), we find
\begin{equation}
\Delta^2_{\beta}=\tfrac{1}{12}\beta\int_0^1(1-\lambda)^{\beta+1}\,d\lambda=\begin{cases}
1/36&\text{for}\;\;\beta=1,\\
1/24&\text{for}\;\;\beta=2.
\end{cases}
\end{equation}

\section{Comparison with voltage-probe model}
\label{sec_comparison}

To compare with the voltage-probe model of dephasing \cite{But86} we work in the framework of RMT, where we have closed form expressions for the transferred charge. We consider the multi-mode regime of $M\gg 1$ monitoring modes in the closed lead, all with the same measurement strength $w^2=\gamma/M\ll 1$.

Generalizing the previous section, we allow for an arbitrary number $N=N_1+N_2$ of modes in the open lead. We seek the charge ${\cal T}$ transferred from contact 1 ($N_1$ modes) into contact 2 ($N_2$ modes). Eq.\ \eqref{conjecture2} generalizes to
\begin{equation}
{\cal T}=\sum_{n=1}^{N_1}\left(\frac{N_2}{N}\sum_{m=N+1}^{\cal N}|\tilde{S}_{mn}|^2+\sum_{n'=N_1+1}^N|\tilde{S}_{n'n}|^2\right).
\label{conjecture3}
\end{equation}
The ${\cal N}\times {\cal N}$ matrix $\tilde{S}$, constructed in Eq.\ \eqref{tildeSdef}, has the Poisson kernel distribution \cite{Bee97} inherited from the circular ensemble for $S$ (the CUE for $\beta=2$, the COE for $\beta=1$).

It is helpful to partition $\tilde{S}$ into blocks, referring to the partition ${\cal N}=N_1+N_2+M$:
\begin{equation}
\tilde{S}=\begin{pmatrix}
s_{11}&s_{12}&s_{13}\\
s_{21}&s_{22}&s_{23}\\
s_{31}&s_{32}&s_{33}
\end{pmatrix}.
\end{equation}
A submatrix $s_{ij}$ is of size $N_i\times N_j$ with $N_3\equiv M$. We define the scalar
\begin{equation}
G_{kl}=\delta_{kl}N_k-\operatorname{Tr} s_{kl}^{\vphantom{\dagger}}s_{kl}^\dagger.
\end{equation}
Unitarity of $\tilde{S}$ then allows to rewrite Eq.\ \eqref{conjecture3} identically as
\begin{equation}
{\cal T}=\frac{N_2}{N}G_{11}-\frac{N_1}{N}G_{21}.\label{calTmonitoring}
\end{equation}

Using the same notation, the transferred charge in the voltage-probe model is \cite{Bro97}
\begin{equation}
{\cal T}_{\rm VP}=\frac{G_{11}G_{22}-G_{12}G_{21}}{G_{11}+G_{12}+G_{21}+G_{22}}.\label{TVPdef}
\end{equation}
This expression is obtained by computing the current that leaves the system via modes $n=N+1,N+2,\ldots N+M$, and then reinjecting charge uniformly and incoherently into these $M$ modes, so that no net current is drawn \cite{But86}. 

\begin{figure}[tb]
\centerline{\includegraphics[width=0.7\linewidth]{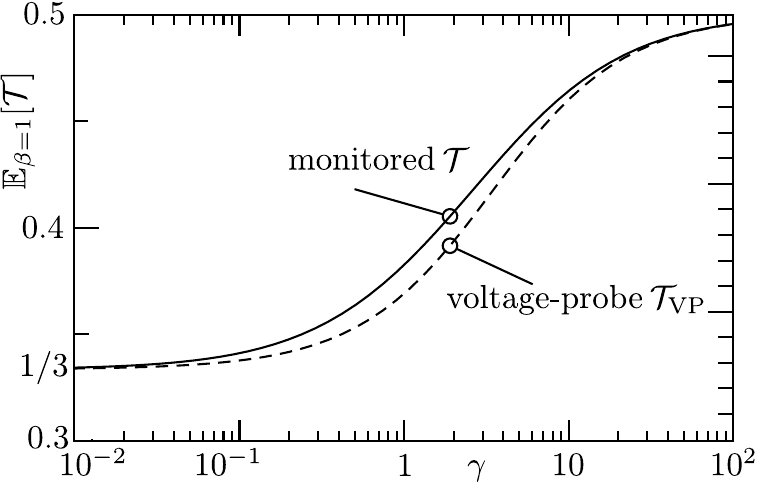}}
\caption{Mean transferred charge for $\beta=1$, computed by averaging with the Poisson kernel of Eqs.\ \eqref{calTmonitoring} (solid curve) and \eqref{TVPdef} (dashed curve). The integral expressions for these averages are given in App. \ref{app_meancharge}.
}
\label{fig_comparison}
\end{figure}

We compare the two models in Fig.\ \ref{fig_comparison} where the $\gamma$ dependence of the mean transferred charge for $\beta=1$ is compared. The results are close. The main qualitative difference appears when we consider the reciprocity of the measurement: The voltage probe transmission \eqref{TVPdef} remains the same if we exchange contacts 1 and 2, while the monitored result \eqref{calTmonitoring} does not.

We emphasise that both models are charge-conserving, the difference is that monitoring exchanges energy with the environment, while the voltage probe does not. The difference appear because of the time scale of the dephasing process: Monitoring is a time-resolved projective measurement, on a time scale $\delta t= L/v_{\rm F}$ much less than the mean time $\tau_{\rm dwell}= (M/N)L/v_{\rm F}$ that an electron stays in the quantum dot (of linear dimension $L$). The voltage-probe model operates on the time scale $\tau_V=\hbar/eV$, set by the energy spread of the conduction electrons. In the low-voltage, linear response regime, $\tau_V\gg\tau_{\rm dwell}\gg\delta t$. Then the voltage probe is a quasi-elastic scatterer, while monitoring is an inelastic process.

\section{Shot noise}
\label{sec_noise}

So far we considered only the first moment of the charge operator $\hat{Q}$, which is the quantity that determines the differential conductance. Higher moments are encoded in the generating function
\begin{equation}
F(\xi)=\lim_{t\rightarrow\infty}\operatorname{Tr}e^{\xi\hat{Q}}\rho(t).
\end{equation}
We may evaluate this in the case $N=2$ of single-mode source and drain contacts, when the density matrix is fully in the single-particle sector of Fock space.

We substitute
\begin{align}
e^{\xi\hat{Q}}={}& \exp\left(\xi\sum_{x=1}^\infty a^{\rm out}_2(x)^\dagger a^{\rm out}_2(x)^{\vphantom{\dagger}}\right)\nonumber\\
={}& \prod_{x=1}^\infty\bigl[1+(e^\xi-1)a^{\rm out}_2(x)^\dagger a^{\rm out}_2(x)^{\vphantom{\dagger}}\bigr]
\end{align}
into the generating function, which reduces to
\begin{equation}
\begin{split}
&F(\xi)=1+(e^\xi-1){\cal T},\\
&{\cal T}=\lim_{t\rightarrow\infty}\sum_{x=1}^\infty \operatorname{Tr}a^{\rm out}_2(x)^\dagger a^{\rm out}_2(x)^{\vphantom{\dagger}}\rho(t),
\end{split}\label{Fxibinomial}
\end{equation}
because $a^{\rm out}_2(x)a^{\rm out}_2(x')\rho(t)=0$ in the single-particle sector. Eq.\ \eqref{Fxibinomial} expresses binomial statistics of the transferred charge for a single injected particle, with transmission probability ${\cal T}$. This is a general property of fermionic statistics \cite{Lev96}, preserved by monitoring \cite{Bee25}.

If the source is biased at a voltage $V$, the total number of particles entering the system in that energy interval during a counting time $t_{\rm counting}$ is ${\cal N}_V=eVt_{\rm counting}/h$. For binomial statistics, the mean and variance of the transferred charge are given by
\begin{equation}
\overline{Q}={\cal N}_V{\cal T},\;\;\operatorname{Var} Q={\cal N}_V {\cal T}(1-{\cal T}),
\end{equation}
corresponding to differential conductance and shot noise power
\begin{equation}
G=(e^2/h){\cal T},\;\;P_{\rm noise}=(e^3/h){\cal T}(1-{\cal T}).
\end{equation}

This is for asymmetric bias, one contact at $V$, the other contact at zero voltage. If instead we bias symmetrically, one contact at $V/2$, the other contact at $-V/2$, we have two independent charge transfer processes with total variance
\begin{align}
\operatorname{Var} Q={}&{\cal N}_V \left[\tfrac{1}{2}{\cal T}(1-{\cal T})+\tfrac{1}{2}{\cal T}'(1-{\cal T}')\right]\nonumber\\
={}&{\cal N}_V\left[\overline{\cal T}(1-\overline{\cal T})-\tfrac{1}{4}({\cal T}-{\cal T}')^2\right].
\end{align}
We see that the reciprocity breaking quantifier $\Delta^2_\beta$ computed in Sec.\ \ref{sec_reciprocity} determines how much the shot noise power for symmetric bias is reduced below the binomial value.

\section{Conclusion}
\label{sec_conclude}

Our main purpose in this study was methodological: Can we generalize the chaotic scattering ensemble introduced by Bl\"{u}mel and Smilansky \cite{Blu90} to include projective measurements? The ensemble of Kraus operators that we constructed by alternating unitaries and projectors serves that role, in the context of monitored quantum transport. The resulting quantum master equation for the single-particle density matrix can be solved efficiently, since the sum over measurement outcomes is carried out algebraically. We have made contact with a random-matrix description, which allows for insight and closed-form results. The RMT connection is heuristic, it relies on an ``equipartition rule'' that we have not proven, but which is supported by the close agreement with the master equation. 

The construction leads to several concrete consequences. For single-mode and multi-mode monitoring we obtained the full probability distribution of the differential conductance, both for broken and preserved time-reversal symmetry. The monitoring-induced breaking of reciprocity distinguishes asymmetric and symmetric voltage biasing. In the multi-mode weak-measurement regime the RMT description reduces the problem to Poisson-kernel averages, allowing a direct comparison with B\"uttiker's voltage-probe model.

Beyond the methodological development, our study may have implications for experiments where the monitoring is effectively performed by a high-frequency electromagnetic environment. Time-resolved monitoring requires a measurement time $\delta t$ shorter than the dwell time $\tau_{\rm dwell}$ of electrons in the quantum dot. With single-channel contacts $\tau_{\rm dwell}\simeq\hbar/\delta E$ is of the order of the inverse mean level spacing $\delta E$. For $\delta E\simeq 1\,\text{meV}$ this corresponds to time scales in the sub-picosecond range. Electromagnetic noise (gate voltage fluctuations) in semiconductor quantum dots typically has a much longer correlation time, but high-frequency fluctuations can be deliberately engineered and investigated using on-chip antennas or microwave resonators.

A natural mathematical problem left open by this work is to prove the equipartition rule as a limiting theorem for the circular ensemble, as specified in Conjectures \eqref{conjecture} and \eqref{conjecture2}.

\acknowledgments

This work was supported by the Netherlands Organisation for Scientific Research (NWO/OCW), as part of Quantum Limits (project number {\sc summit}.1.1016).\\
Computer codes and data sets are available at a \href{https://doi.org/10.5281/zenodo.20523817}{Zenodo repository}.

\appendix

\section{From sum over quantum trajectories to master equation}
\label{app_mastereq}

Following the general approach of Ref.\ \onlinecite{San26}, we work out the algebra that transforms the sum \eqref{rhotsingleparticle} over quantum trajectories into the master equation \eqref{mastereq}.

We rewrite Eq.\ \eqref{rhotsingleparticle} as
\begin{equation}
\begin{split}
&{\rho}(t+1)=\sum_{\bm s}{\cal P}_{\bm s}\tilde{\rho}(t){\cal P}_{\bm s},\\
&{\cal P}_{\bm s}= \prod_{n}{P}_{s_n,n},\;\;\tilde{\rho}(t)=DS\rho(t)(DS)^\dagger.
\end{split}\label{calPdef}
\end{equation}
We include all modes in the product, in both open and closed lead, which does not change the expression if we set the measurement probability $w_n=0$ for modes in the open lead.

The binary string $\bm{s}$ selects a matrix ${\cal P}_{s_n,n}$ of the form
\begin{equation}
\begin{split}
&{\cal P}_{s_n,n}= a_{s_n,n} P_n +b_{s_n,n} Q_n,\\
&P_n=|n\rangle\langle n|=I-Q_n,\\
&a_{\pm,n}=2^{-1/2}\sqrt{1\pm w_n},\;\;b_{\pm,n}=2^{-1/2}\sqrt{1\mp w_n}.
\end{split}\label{Asmmdef}
\end{equation}
The normalization
\begin{equation}
a_{+,n}^2+a_{-,n}^2=1=b_{+,n}^2+b_{-,n}^2\label{abnormalize}
\end{equation}
ensures the sum rule
\begin{equation}
\sum_{\bm s}{\cal P}_{\bm s} {\cal P}^{\vphantom{\dagger}}_{\bm s}=I.
\end{equation}

Substitution of Eq.\ \eqref{Asmmdef} into Eq.\ \eqref{calPdef} produces a sum of products of $P$ and $Q$ projectors. Each product cannot contain more than one $P$ projector, because the projectors commute and $P_n P_m=0$ if $n\neq m$. Moreover, $\prod_{n} Q_n =0$, so each product must contain precisely one $P$ projector, say $P_n$. Summation over the products then gives
\begin{align}
{\cal P}_{\bm s}={}& \sum_{n=1} a _{s_n,n}P_n\prod_{m\neq n}b_{s_m,m}Q_m\nonumber\\
={}&\sum_{n=1} a_{s_n,n}P_n\prod_{m\neq n}b_{s_m,m},
\end{align}
where in the second equality we used that $P_n Q_m=P_n$ for $n\neq m$. We thus have
\begin{widetext}
\begin{equation}
\sum_{\bm s}{\cal P}_{\bm s}\tilde{\rho}(t){\cal P}_{\bm s}=\sum_{\bm s}\biggl(\sum_{n} a_{s_n,n}P_n\prod_{i\neq n}b_{s_i,i}\biggr)\tilde{\rho}(t)\biggl(\sum_{m} a_{s_m,m}P_m\prod_{j\neq m}b_{s_j,j}\biggr).
\end{equation}
\end{widetext}

The sum over the strings $\bm{s}$ can be worked out further by splitting the double sum over the mode indices into $n=m$ and $n\neq m$,
\begin{equation}
\begin{split}
&\sum_{\bm s}{\cal P}_{\bm s}\tilde{\rho}(t){\cal P}_{\bm s}=\sum_{n\neq m}c_nc_mP_n\tilde{\rho}(t)P_m+\sum_{n}P_n\tilde{\rho}(t)P_n,\\
&c_n=\sum_{s_n} a_{s_n,n}b_{s_n,n}=\sqrt{1-w_n^2},
\end{split}
\end{equation}
where we have used the normalization \eqref{abnormalize}. One more step, to remove the restriction $n\neq m$,
\begin{align}
\sum_{\bm s}{\cal P}_{\bm s}\tilde{\rho}(t){\cal P}_{\bm s}={}&\sum_{n,m}c_n c_m|n\rangle\langle m|\tilde{\rho}_{nm}(t)\nonumber\\
&+\sum_{n}w_n^2|n\rangle\langle n|\tilde{\rho}_{nn}(t),
\end{align}
with $\tilde{\rho}_{nm}=\langle n|\tilde{\rho}|m\rangle$. Jointly with Eq.\ \eqref{calPdef} we arrive at the master equation \eqref{mastereq} from the main text (noting that we set $w_n=0$ for $n$ not in the closed lead).

\section{Lyapunov method of  solution of the master equation}
\label{app_Lyapunov}

The Lyapunov method allows for an efficient solution of the quantum master equation for the matrix $\Xi=\sum_{t=0}^\infty\rho(t)$, which determines the transferred charge ${\cal T}$. We explain this first for the case of a single monitored mode considered in Sec.\ \ref{sec_singlemode} \cite{note5}.

\subsection{Single-mode case}

Upon summation of Eq.\ \eqref{monitoredmastereq} over $t$ we have
\begin{equation}
    \Xi = Q DS \,\Xi (QDS)^\dagger  + P DS\,\Xi (PDS)^\dagger  + \rho(0).\label{Ximonitored}
\end{equation}
Given that $\rho(0)={\cal S}\chi_{\rm in}$ has no overlap with the monitored mode, $P\rho(0)=0=\rho(0)P$, the matrix $\Xi$ has only components in the $P$ and $Q$ subspaces,
\begin{equation}
\Xi=\Xi_{QQ} + \chi |3\rangle\langle 3|,\;\;\Xi_{QQ}=Q\Xi Q.
\end{equation}
 
We project the master equation onto the $Q$-subspace,
\begin{equation}
    \Xi_{QQ} = Q DS(\Xi_{QQ} + \chi |3\rangle\langle 3| ) (QDS)^\dagger + \rho(0).
\end{equation}
We define the projected operator ${\cal V} = QDSQ$ and the rank-one matrix
\begin{equation}
    {\cal M}= QDS |3\rangle\langle 3| (QDS)^\dagger .
\end{equation}
The equation for $\Xi_{QQ}$ then takes the form of a discrete Lyapunov equation (also known as Stein equation) \cite{Sim16},
\begin{equation}
    \Xi_{QQ} - {\cal V}\Xi_{QQ} {\cal V}^\dagger = \chi {\cal M} + \rho(0).
\end{equation}
This equation can be solved with ${\cal O}({\cal N}^3)$ complexity by Schur decomposition of ${\cal V}$ \cite{Bar72,Kit77}.

To determine the scalar $\chi$ we project Eq.\ \eqref{Ximonitored} onto the $P$-subspace,
\begin{equation}
    \chi = \langle 3| DS (  \Xi_{QQ} +\chi |3\rangle\langle 3|) (DS)^\dagger |3\rangle .
\end{equation}
Defining $\alpha = |\langle 3| DS |3\rangle|^2$ this may be rearranged as
\begin{equation}
    \chi(1 - \alpha) = \langle 3|DS \Xi_{QQ} (DS)^\dagger |3\rangle.
\end{equation}

The calculation of ${\cal T}$ then proceeds as follows. 
\begin{itemize}
\item Decompose $\Xi_{QQ} = \chi \Xi_1 + \Xi_2$ and use the linearity of the Lyapunov equation to solve separately for $\Xi_1$ and $\Xi_2$ from
\[
\Xi_1 - {\cal V} \Xi_1 {\cal V}^\dagger = {\cal M},\;\;\Xi_2 - {\cal V} \Xi_2 {\cal V}^\dagger = \rho(0).
\]
    \item Substitute $\Xi_{QQ}$ into the scalar equation and solve for $\chi$,
\[
        \chi= \frac{\langle 3| DS \Xi_2 (DS)^\dagger |3\rangle}{1 - \alpha - \langle 3| DS\Xi_1 (DS)^\dagger |3\rangle }.
\]
    \item Construct the full solution,
    \[
    \Xi = \chi |3\rangle\langle 3| + \chi \Xi_1 + \Xi_2 ,
    \]
    and compute ${\cal T}=\operatorname{Re}\operatorname{Tr}\chi_{\rm out} \Xi$ from Eq.\ \eqref{calTsinglemode},
\end{itemize}

\subsection{Multi-mode case}

The ${\cal O}({\cal N}^3)$ complexity is for a single monitored mode. In the more general $M$-mode case of Sec.\ \ref{sec_multimode}, with master equation \eqref{mastereq}, we need to solve a set of $M$ coupled linear equations instead of the single equation for $\chi$. The complexity of the algorithm then increases to ${\cal O}({\cal N}^4)$, which is still more efficient than the ${\cal O}({\cal N}^6)$ solution by vectorization. We outline the approach for that case.

Summation of Eq.\ \eqref{mastereq} over $t$ gives for $\Xi=\sum_{t=0}^\infty\rho(t)$ the equation
\begin{align}
&\Xi=\rho(0)+QDS\,\Xi(QDS)^\dagger\nonumber\\
&\quad+\sum_{m=N+1}^{\cal N}w_m^2|m\rangle\langle m|DS\,\Xi(DS)^\dagger|m\rangle\langle m|,
\end{align}
which is a rank-${\cal N}$ perturbation of the discrete Lyapunov equation. The action of the linear operator
\begin{equation}
{\cal A}(\Xi)=\Xi-QDS\,\Xi(QDS)^\dagger,\label{Sdef}
\end{equation} can be inverted in order ${\cal N}^3$ complexity.
The equation for $\Xi$ can be rewritten as
\begin{align}
&{\cal A}(X)=\rho_0+\sum_{m=N+1}^{\cal N}w_m^2 x_{m}P_m,\\
&P_m=|m\rangle\langle m|,\;\;x_m=\langle m|DS\,\Xi(DS)^\dagger|m\rangle.
\end{align}

We now first solve for $\Xi$ in terms of the $x_{m}$ coefficients,
\begin{equation}
\Xi={\cal A}^{-1}(\rho_0)+\sum_{m=N+1}^{\cal N}w_m^2 x_{m}{\cal A}^{-1}(P_m).
\end{equation}
We next conjugate this equation with $DS$ and project onto mode $m$, to obtain a set of coupled equations for the $x_{m}$ coefficients,
\begin{align}
&x_{m}=\langle m|DS{\cal A}^{-1}(\rho_0)(DS)^{\dagger}|m\rangle\nonumber\\
&\;+\sum_{m'=N+1}^{\cal N}w_{m'}^2 x_{m'}\langle m|DS{\cal A}^{-1}(P_{m'})(DS)^{\dagger}|m\rangle.\label{xmdef}
\end{align}
To compute this efficiently, we use that, by definition \eqref{Sdef}, 
\begin{equation}
{\cal A}^{-1}(P_{m'}) = \sum_{k=0}^\infty (QDS)^k P_{m'} ((QDS)^\dagger)^k.
\end{equation}
We consider $\widetilde{QDS}$, the truncation of $QDS$ acting only on the $M$ closed modes. Its eigenvalue/eigenvector decomposition is
\begin{equation}
    \widetilde{QDS} = V{\cal E} V^{-1} ,\;\;{\cal E}=\mathrm{diag}(E_{N+1}, \dots E_{\cal N}).
\end{equation}
The coefficient appearing in Eq.\ \eqref{xmdef} can then be computed from
\begin{align}
 &   \langle m|DS{\cal A}^{-1}(P_{m'})(DS)^{\dagger}|m\rangle =\nonumber\\ &=\sum_{a,b = N+1}^{\cal N}\frac{ (DSV)_{ma}(DSV)^*_{mb}
     (V^{-1})_{am'} (V^{-1})^*_{bm'} }{1-E_a^{\vphantom{*}} E_b^*}.
\end{align}

The set of equations \eqref{xmdef} is constructed in order ${\cal N}^4$ complexity and solved in order ${\cal N}^3$ complexity, so the total algorithm has order ${\cal N}^4$ complexity.

\section{Evaluation of the CUE and COE averages}
\label{app_CUECOE}

These are the calculations for Secs.\ \ref{singlemodeRMTCUE} and \ref{singlemodeRMTCOE}

\subsection{Case of asymmetric bias}

Marginal distributions of the elements of matrices in the CUE or COE have been calculated in Refs.\ \onlinecite{Per83,Fri85,Bro95}. Specifying those general results to  $3\times 3$ matrices $U$ we find that
\begin{align}
\begin{rcases}
P_{\rm CUE}(\tau_{21},\tau_{31})=2\\
P_{\rm COE}(\tau_{21},\tau_{31})=(\tau_{21}+\tau_{31})^{-1}
\end{rcases}
\text{if}\;\;\tau_{21}+\tau_{31}<1,
\end{align}
where we have defined $\tau_{nm}=|U_{nm}|^2\in(0,1)$. This gives the probability density functions \eqref{PTCUE} and \eqref{PTCOE} for ${\cal T}=\tau_{21}+\tfrac{1}{2}\tau_{31}$.

\subsection{Case of symmetric bias}

For the distribution of
\begin{align}
\overline{\cal T}={}&\tfrac{1}{2}\tau_{21}+\tfrac{1}{4}\tau_{31}+\tfrac{1}{2}\tau_{12}+\tfrac{1}{4}\tau_{32}\nonumber\\
={}&\tfrac{1}{4}\bigl(2+\tau_{12}+\tau_{21}-\tau_{11}-\tau_{22}\bigr)\nonumber\\
={}&\tfrac{1}{2}-\tfrac{1}{4}\operatorname{Tr}\bm{u}\sigma_z\bm{u}^\dagger\sigma_z,\label{calTu}
\end{align}
with Pauli matrix $\sigma_z$, we need the marginal distribution of the $2\times 2$ upper-left submatrix $\bm{u}$ of $U$. This follows from the general formulas in Ref.\ \onlinecite{Bee97}.

The submatrix $\bm{u}$ has the singular value decomposition
\begin{equation}
\bm{u}=V_1\Lambda V_2,\;\;\Lambda=\begin{pmatrix}
1&0\\
0&\sqrt\lambda
\end{pmatrix},
\end{equation}
with $2\times 2$ unitary matrices $V_1,V_2$ and real $\lambda\in(0,1)$. Unitarity of $U$ pins one of the singular values at unity, while the other has probability density function \cite{Bee97}
\begin{equation}
P(\lambda)=\beta(1-\lambda)^{\beta-1},
\end{equation}
with $\beta=1$ in the COE and $\beta=2$ in the CUE.

In the CUE the matrices $V_1$ and $V_2$ have independent Haar-uniform distributions in ${\rm U}(2)$, in the COE $V_2=V_1^\top$ is Haar-uniform in ${\rm U}(2)$. Substitution into Eq.\ \eqref{calTu} gives
\begin{equation}
\overline{\cal T}=\tfrac{1}{2}-\tfrac{1}{4}\operatorname{Tr}\Lambda (V_2^{\vphantom{\dagger}}\sigma_z V_2^\dagger)\Lambda (V_1^\dagger\sigma_zV_1^{\vphantom{\dagger}}),
\end{equation}
which may equivalently be written as
\begin{equation}
\overline{\cal T}=\tfrac{1}{2}-\tfrac{1}{4}\operatorname{Tr}\Lambda (\bm{n}_2\cdot\bm{\sigma})\Lambda (\bm{n}_1\cdot\bm{\sigma}).
\end{equation}
The unit vectors $\bm{n}_1,\bm{n}_2\in\mathbb{R}^3$ are independently and uniformly distributed on the unit sphere for the CUE, while for the COE one has 
\begin{equation}
\overline{\cal T}=\tfrac{1}{2}-\tfrac{1}{4}\operatorname{Tr}\Lambda (\bm{n}\cdot\bm{\sigma})\Lambda (\bm{n}\cdot\bm{\sigma})^\top\nonumber\\
\end{equation}
with $\bm{n}$ uniform on the unit sphere.

For the CUE we define $\overline{\cal T}=\tfrac{1}{2}(1-X)$ with
\begin{equation}
X=\sqrt{\lambda}(n_{1,x}n_{2,x}+n_{1,y}n_{2,y})+\tfrac{1}{2}(1+\lambda)n_{1,z}n_{2,z}.
\end{equation} 
The probability density function of $X$ conditioned on $\lambda$, determined by the independent uniformly distributed unit vectors $\bm{n}_1$ and $\bm{n}_2$, is
\begin{widetext}
\begin{equation}
P(X|\lambda)=
\begin{cases}
(1-\lambda)^{-1}\ln(1/\sqrt{\lambda}),
&
|X|< \sqrt{\lambda},
\\
(1-\lambda)^{-1}[\ln(1/\sqrt{\lambda})-\operatorname{arcosh}(|X|/\sqrt{\lambda})],
&
\sqrt{\lambda}< |X|< \tfrac{1}{2}(1+\lambda),
\\
0,
&
|X|>\tfrac{1}{2}(1+\lambda).
\end{cases}
\end{equation}
\end{widetext}
We then average over $\lambda\in(0,1)$, with density $P(\lambda)=2(1-\lambda)$,
\begin{equation}
P(X)=
\begin{cases}
1-2X^2,
&
|X|< \tfrac{1}{2},
\\
2(1-|X|)^2,
&
\tfrac{1}{2}<|X|< 1,
\\
0,
&
|X|>1,
\end{cases}
\end{equation}
which gives Eq.\ \eqref{PTbarCUE}.

For the COE we define $\overline{\cal T}=\tfrac{1}{2}(1-Y)$ with
\begin{equation}
Y=\sqrt{\lambda}(n_{x}^2-n_y^2)+\tfrac{1}{2}(1+\lambda)n_z^2.
\end{equation}
We condition on both $\lambda$ and $n_z$,
\begin{equation}
\begin{split}
&P(Y|\lambda,\zeta)=\frac{\theta(\omega)}{\pi\sqrt\omega},\\
&\omega=
\lambda(1-n_z^2)^2-\left[Y-\tfrac{1}{2}(1+\lambda)n_z^2\right]^2,
\end{split}
\end{equation}
with $\theta(\omega)$ the unit step function. Averaging over the uniform distributions of $\lambda\in(0,1)$ and $n_z\in(-1,1)$ results in
\begin{equation}
P(Y)=
\begin{cases}
2-\sqrt{2(1-Y)}, & -1< Y< \tfrac{1}{2},\\
\sqrt{2(1-Y)}, & \tfrac{1}{2}< Y< 1,\\
0, & |Y|>1,
\end{cases}
\end{equation}
which amounts to Eq.\ \eqref{PTbarCOE}.

\section{Evaluation of the Poisson kernel averages}
\label{app_Poisson}

These are the calculations for Secs.\ \ref{multimodeRMT} and \ref{sec_reciprocity}.

\subsection{Case of asymmetric bias}

We use unitarity of $\tilde{S}$ to rewrite Eq.\ \eqref{conjecture2} identically as
\begin{align}
{\cal T}={}&\tfrac{1}{2}+\tfrac{1}{2}|\tilde{S}_{21}|^2-\tfrac{1}{2}|\tilde{S}_{11}|^2\nonumber\\
={}&\tfrac{1}{2}-\tfrac{1}{4}\operatorname{Tr}\tilde{\bm u}^\dagger\sigma_z \tilde{\bm u}(1+\sigma_z),
\end{align}
with $\tilde{\bm u}$ the $2\times 2$ upper-left block of the ${\cal N}\times{\cal N}$ unitary matrix $\tilde{S}$ defined in Eq.\ \eqref{tildeSdef}. The matrix $\tilde{S}$ has the Poisson kernel distribution inherited from the circular ensemble for $S$ (the CUE for $\beta=2$, the COE for $\beta=1$).

The matrix $\tilde{\bm u}$ has the singular value decomposition
\begin{equation}
\tilde{\bm u}=V_1\tilde{\Lambda }V_2,\;\;\tilde{\Lambda}=\begin{pmatrix}
\sqrt{\lambda_1}&0\\
0&\sqrt{\lambda_2}
\end{pmatrix}.
\end{equation}
The unitary matrices $V_1,V_2$ are independently Haar uniform in ${\rm U}(2)$ for $\beta=2$, while for $\beta=1$ $V_2=V_1^\top$ is Haar uniform in ${\rm U}(2)$. The singular values are independent of the unitaries. The probability density function $P_\beta(\lambda_1,\lambda_2)$ of $\lambda_1,\lambda_2\in(0,1)$ is computed from the Poisson kernel in Ref.\ \onlinecite{Bro97}, in the limit $M\rightarrow\infty$, $w\rightarrow 0$ at fixed $\gamma=Mw^2$. 

For $\beta=2$ we first average over the unitaries $V_1,V_2$ to obtain the conditional distribution $P({\cal T}|\lambda_1,\lambda_2)$ of ${\cal T}=\tfrac{1}{2}(1-X)$ for given $\lambda_1,\lambda_2$,
\begin{widetext}
\begin{equation}
P_{\beta=2}({\cal T}\mid \lambda_1,\lambda_2)=
\begin{cases}
\displaystyle
\frac{2}{|\lambda_1-\lambda_2|}
\operatorname{arsinh}\!\left(\frac{|\lambda_1-\lambda_2|}{2\sqrt{\lambda_1\lambda_2}}\right),
&
|X|< \lambda_{\rm min},
\\
\displaystyle
\frac{1}{|\lambda_1-\lambda_2|}
\left[
\operatorname{arsinh}\!\left(\frac{|\lambda_1-\lambda_2|}{2\sqrt{\lambda_1\lambda_2}}\right)
-
\operatorname{arsinh}\!\left(
\frac{X^2-\lambda_1\lambda_2}{2|X|\sqrt{\lambda_1\lambda_2}}
\right)
\right],
&
\lambda_{\rm min}<|X|<\lambda_{\rm max},
\\
0,
&
|X|> \lambda_{\rm max},
\end{cases}
\end{equation}
\end{widetext}
with $\lambda_{\rm min}=\min(\lambda_1,\lambda_2)$, $\lambda_{\rm max}=\max(\lambda_1,\lambda_2)$. Subsequently we numerically evaluate \cite{note4}
\begin{equation}
P_\beta({\cal T})=\int_0^1 d\lambda_1\int_0^1 d\lambda_2 \,P_\beta({\cal T}|\lambda_1,\lambda_2)P_\beta(\lambda_1,\lambda_2).\label{Pbetaeq}
\end{equation}

For $\beta=1$ there is a single unitary to average over, which gives the elliptic integral
\begin{subequations}
\begin{align}
&P_{\beta=1}({\cal T}|\lambda_1,\lambda_2)=\int_0^1 dz\,\frac{\theta(\omega')}{\pi\sqrt{\omega'}},\\
&\omega'=4\lambda_1\lambda_2 z^2(1-z)^2\nonumber\\
&\qquad-\bigl({\cal T}-\tfrac{1}{2}-(\tfrac{1}{2}-z)[\lambda_1 z -\lambda_2(1-z)]\bigr)^2,
\end{align}
\end{subequations}
and hence $P_\beta({\cal T})$ upon substitution into Eq.\ \eqref{Pbetaeq} and numerical integration.

\subsection{Mean transferred charge}
\label{app_meancharge}

The first moment has a closed form expression,
\begin{widetext}
\begin{align}
\mathbb{E}[{\cal T}]_{\beta=1}={}&\tfrac{1}{2}-\tfrac{1}{12}\int_0^1 d\lambda_1\int_0^1 d\lambda_2 \,(\lambda_1+\lambda_2)P_{\beta=1}(\lambda_1,\lambda_2)\nonumber\\
={}&2\gamma^{-1} \left[  \left(2 e^{{\gamma}}-\gamma^2-2\gamma-2\right) \operatorname{Ei}(-{\gamma})- \left( \tfrac{1}{2}{\gamma}^2- {\gamma}+1-e^{-{\gamma}}\right) e^{{\gamma}/2}\operatorname{Ei}\left(-\gamma/2\right)+1-(1+\gamma)e^{-\gamma}\right],
\end{align}
with Ei the exponential integral function. This is the solid curve in Fig.\ \ref{fig_comparison}.

For comparison also the mean transferred charge in the voltage-probe model is plotted in Fig.\ \ref{fig_comparison}. That first moment follows from Eq.\ \eqref{TVPdef},
\begin{equation}
\mathbb{E}[{\cal T}_{\rm VP}]_{\beta=1}=\tfrac{1}{3}+\tfrac{1}{3}\int_0^1 d\lambda_1\int_0^1 d\lambda_2 \,\frac{(1-\lambda_1)(1-\lambda_2)}{2-\lambda_1-\lambda_2}P_{\beta=1}(\lambda_1,\lambda_2).
\end{equation}
\end{widetext}
We have evaluated this integral numerically (a closed-form expression was not forthcoming).

\subsection{Case of symmetric bias}

Turning next to the symmetric bias, we have the transferred charge
\begin{equation}
\overline{\cal T}=\tfrac{1}{2}({\cal T}+{\cal T}')=\tfrac{1}{2}-\tfrac{1}{4}\operatorname{Tr}\tilde{\bm{u}}\sigma_z\tilde{\bm{u}}^\dagger\sigma_z.\label{overlinecalTdef}
\end{equation}

The conditional probability density function for $\beta=2$ now takes the form
\begin{widetext}
\begin{equation}
P_{\beta=2}(\overline{\cal T}\mid \lambda_1,\lambda_2)=
\begin{cases}
\displaystyle
\frac{2}{|\lambda_1-\lambda_2|}
\operatorname{arsinh}\!\left(\frac{|\lambda_1-\lambda_2|}{2\sqrt{\lambda_1\lambda_2}}\right),
&
|Y|< \sqrt{\lambda_1\lambda_2},
\\
\displaystyle
\frac{2}{|\lambda_1-\lambda_2|}
\left[
\operatorname{arsinh}\!\left(\frac{|\lambda_1-\lambda_2|}{2\sqrt{\lambda_1\lambda_2}}\right)
-
\operatorname{arsinh}\!\left(
\frac{\sqrt{Y^2-\lambda_1\lambda_2}}{\sqrt{\lambda_1\lambda_2}}
\right)
\right],
&
\sqrt{\lambda_1\lambda_2}<|Y|<\tfrac{1}{2}(\lambda_1+\lambda_2),
\\
0,
&
|Y|> \tfrac{1}{2}(\lambda_1+\lambda_2),
\end{cases}
\end{equation}
\end{widetext}
with $\overline{\cal T}=\tfrac{1}{2}(1-Y)$.

For $\beta=1$ we again have an elliptic integral,
\begin{subequations}
\begin{align}
&P_{\beta=1}(\overline{\cal T}|\lambda_1,\lambda_2)=\int_0^1 dz\,\frac{\theta(\omega'')}{\pi\sqrt{\omega''}},\\
&\omega''=4\lambda_1\lambda_2 z^2(1-z)^2\nonumber\\
&\qquad-\bigl({\cal T}-\tfrac{1}{2}-(\lambda_1+\lambda_2)[z(1-z)-\tfrac{1}{4}]\bigr)^2.
\end{align}
\end{subequations}
We integrate these conditional distributions numerically, weighted by $P_\beta(\lambda_1,\lambda_2)$, to obtain the curves in Fig.\ \ref{fig_histograms}.

\subsection{Reciprocity breaking}

The difference ${\cal T}-{\cal T}'$ of the transferred charge from contact 1 to contact 2 and the other way around is given by
\begin{align}
{\cal T}-{\cal T}'={}&-\tfrac{1}{2}\operatorname{Tr}\tilde{\bm{u}}\sigma_z\tilde{\bm{u}}^\dagger\nonumber\\
={}&-\tfrac{1}{2}\operatorname{Tr}\tilde{\Lambda}^2(\bm{n}\cdot\bm{\sigma})=\tfrac{1}{2}(\lambda_2-\lambda_1)n_z,
\end{align}
 with $\bm{n}$ uniformly distributed on the unit sphere.
 
The average vanishes, while the variance is given by
\begin{align}
\Delta^2_\beta={}&\mathbb{E}[({\cal T}-{\cal T}')^2]\nonumber\\
={}&\tfrac{1}{12}\int_0^1 d\lambda_1\int_0^1 d\lambda_2 \,(\lambda_2-\lambda_1)^2P_{\beta}(\lambda_1,\lambda_2),
\end{align}
which evaluates to Eq.\ \eqref{Delta2def} in the main text.


\begin{thebibliography}{99}
\bibitem{note1} For sparse matrices vectorization can become comparably efficient to the Lyapunov method of solution. This was the case in our localization study \cite{San26}.
\bibitem{note2} The invariance of the Haar measure for $S\mapsto SS_0$ and $S\mapsto S_0S$, with arbitrary $S_0\in{\rm U}({\cal N})$, implies the corresponding invariance for $U$ if we take $S_0=U_0\oplus I_{M-1\times M-1}$ and arbitrary $U_0\in{\rm U(3)}$.
\bibitem{note5} The method of solution of the master equation described in App.\ \ref{app_Lyapunov} was suggested to us by AI (GPT-5.5). It is based directly on established linear algebra algorithms \cite{Sim16,Bar72,Kit77}, but we have not found a reference for precisely our application.
\bibitem{note4} The expression for $P_\beta(\lambda_1,\lambda_2)$ is very lengthy and not recorded here, we refer to Ref.\ \onlinecite{Bro97}, Eq.\ (17), with notation $\lambda_n\mapsto 1-T_n$.
\end{thebibliography}
\end{document}